\documentclass[]{raa}
\usepackage{graphicx,times}
\usepackage{natbib}
\usepackage{amsmath}
\usepackage{booktabs,longtable}
\usepackage{pdflscape}

\begin{document}

\title{Metallicity and Star Formation Activities of the Interacting System Arp 86 from Observation with MOS on Xinglong 2.16m Telescope}
   \volnopage{Vol.0 (200x) No.0, 000--000}      
   \setcounter{page}{1}          

   \author{Zhi-Min Zhou
      \inst{1}
   \and Hong Wu
      \inst{1}
   \and Lei Huang
      \inst{1}
   \and Hong-Bin Li
      \inst{1}
   \and Zhi-Zhong Zhou
      \inst{1}
	\and Jun-Jun Jia
	  \inst{1}
	\and Man-I Lam
	  \inst{1}
	\and Yi-Nan Zhu
	  \inst{1}
   }
   \institute{Key Laboratory of Optical Astronomy, National Astronomical Observatories, Chinese Academy of Sciences, Beijing 100012, China; {\it zmzhou@bao.ac.cn}}
  \date{Received~~2014 month day; accepted~~2014~~month day}


\abstract{ We present an analysis of the metallicity and star formation activities of H\,{\sc ii} regions in the interacting system Arp 86, based on the first scientific observation of the multi-object spectroscopy on the 2.16m Telescope at Xinglong Observatory. We find that the oxygen abundance gradient in Arp 86 is flatter than that in normal disk galaxies, which confirms that gas inflows caused by tidal forces during encounters can flatten the metallicity distributions in galaxies. The companion galaxy NGC 7752 is currently experiencing a galaxy-wide starburst with higher surface density of star formation rate than the main galaxy NGC 7753, which can be explained that the companion galaxy is more susceptible to the effects of interaction than the primary. We also find that the galaxy 2MASX J23470758+2926531 has similar abundance and star formation properties to NGC 7753, and may be a part of the Arp 86 system.
\keywords{galaxies: abundances --- galaxies: evolution --- galaxies: interactions --- (ISM:) HII regions --- techniques: spectroscopic}
}

   \authorrunning{Zhou et al.}            
   \titlerunning{Metallicity and Star Formation in Arp 86 }  

   \maketitle

\section{Introduction}
Gravitational interactions and mergers play a major role in galaxy evolution, drastically modifying both their morphology and star formation activities \citep{Barnes1992, Hopkins2010}. Detailed investigations of individual interacting galaxies can provide important information about the physics of the encounter, including the enhancement or suppression of star formation (SF), as well as hydrodynamical processes \citep{Struck2003}.

The metallicity distribution and the properties of stellar populations in galaxies are important to understand the physical processes in the formation and evolution of galaxies and have been studied by many authors \citep[e.g.,][]{Wu2005b, Zhou2011}.
Extragalactic H\,{\sc ii} regions represent perfect laboratories for deriving physical properties of gaseous nebulae and stars clusters across the surface of nearby galaxies \citep{Osterbrock06}. Their characteristic emission-line spectra have been extensively used to probe the stellar populations and chemical composition of local star-forming galaxies, which can provide observational tests for the mechanisms of galaxy evolution \citep{Bresolin2012, LiFabio2013}.

Thanks to the multi-object spectroscopy (MOS), which offers the possibility of obtaining spectra of tens to hundreds objects simultaneously. The development of MOS instruments at ground-based telescopes provides an opportunity to derive physical properties of interacting galaxies using optical spectra of the associated ionized gas. 

In the current study, we present spectroscopic observations of an interacting system Arp 86 with the help of MOS, to probe the metallicity, star formation rates(SFRs) and stellar populations of its H\,{\sc ii} regions, in an attempt to study the impact of interacting on the evolution of galaxies. Arp 86 is an interacting galaxy pair resembling M51 system with a redshift of {\it z} $\sim$ 0.0162 from the NASA/IPAC Extragalactic Database (NED). The galaxy pair consists of a grand-design galaxy NGC 7753, and a small companion NGC 7752. NGC 7753 is a SAB galaxy with the radii R$_{25} \sim$ 1 arcmin (corresponding to 20 kpc), which is defined by the isophote at the brightness of 25 mag/arcsec$^2$ in the B-band. A tidal bridge connecting two galaxies is found in optical images and active star formation is lying in two galaxies and the bridge \citep{Laurikainen1993}. The infrared luminosities (8 -- 1000 $\mu$m) of Arp 86 is $10^{11.01} L_\odot$ \citep{Sanders2003}, indicating that it is a luminous infrared galaxy (LIRG) system. Radio continuum observations show a complex distribution of H{\sc i} tails and bridges due to tidal interactions \citep{Sengupta2009}. H{\sc i} maps show that the compact dwarf galaxy 2MASX J23470758+2926531 lying in the southeast of Arp 86 may also be a part of this system \citep{Sengupta2009}.

The layout of this article is as follows: in Section~\ref{sec2} we introduce the selection of target star-forming regions for spectra, together with a description of the observations and data reduction procedures. In Section~\ref{sec3} we present and analyze the observational results, including dust extinction, excitation properties, metallicities, and star formation activities. The results are discussed and interpreted in Section~\ref{sec4}, and then summarized in Section~\ref{sec5}.

\section{Observation and Data}
\label{sec2}
In this section, we first briefly describe the selection of the target H\,{\sc ii} regions in Arp 86. The instrument we used and observations are presented next. Then the data reduction and line flux measurement are introduced. 

\subsection{Sample Selection}
\begin{figure}[!htbp]
\center
\includegraphics[angle=0,width=0.8\textwidth]{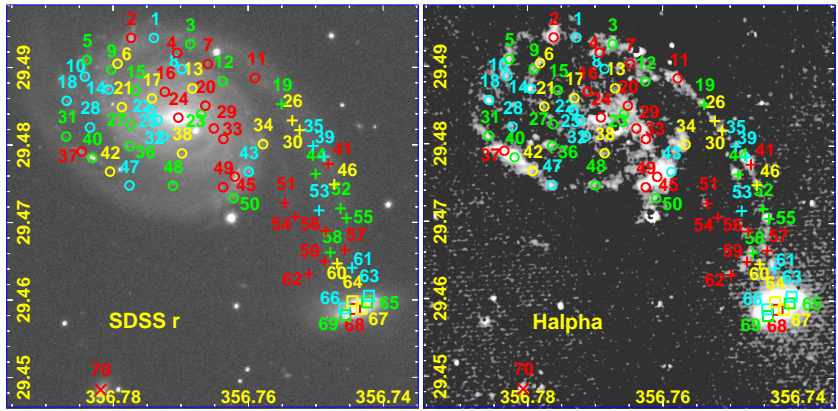}
\caption{images of Arp 86, identifying the target H\,{\sc ii} regions. The background images are SDSS r ({\it left}) and H$\alpha$ ({\it right}) images of Arp 86 with the target H\,{\sc ii} regions labeled. The symbols indicate the different data locations: circles--the dominant galaxy NGC 7753, squares -- the companion galaxy NGC 7752, pluses -- the tidal bridge between the two galaxies, and cross -- the galaxy 2MASX J23470758+2926531. There are four colors and the targets with the same color were observed in the same mask. Region numbers are labeled in the figure and correspond to those given in Table~\ref{table1} 
\label{fig1}}
\end{figure}

We selected the target H\,{\sc ii} regions from a narrowband H$\alpha$ image of Arp 86, which was obtained by the BAO Faint Object Spectrograph and Camera (BFOSC) attached to the 2.16m telescope at Xinglong Observatory of the National Astronomical Observatories of the Chinese Academy of Sciences. The bright H\,{\sc ii} regions were manually selected in the bulge and disks. In order to avoid the selection biases, some faint candidates were also included along with the bright ones. Finally, 69 sources were selected, their spatial locations are shown in Figure~\ref{fig1}. Astronomical coordinates and identification data of these regions are presented in Table~\ref{table1}. Beside the targets in Arp 86, the object 2MASX J23470758+2926531 was also selected (No.70 in Figure~\ref{fig1}).

\subsection{Instrument and Observation}
We obtained optical spectroscopy of H\,{\sc ii} region candidates with the MOS at the f/9 Cassegrain focus of the Xinglong 2.16m telescope. The MOS is attached to the spectroscopic mode of BFOSC, the same instrument used to take the narrowband H$\alpha$ image. The E2V CCD detector that is used with MOS has 22.5 $\mu$m square pixels, yielding an image scale of 0\farcs457 pixel$^{-1}$ and $\sim 8\farcm5 \times 9\farcm5$ field of view (FoV). The MOS instrument can observe ten to one hundred targets simultaneously. 

The new designed MOS is based upon a mask plate with pinholes, which are corresponding to the targets. The mask replaces the slit plate in the spectroscopic system of BFOSC. 
The positions of pinholes are designed according to the astronomical coordinates of the targets and focus parameters of the telescope, and drilled with an precise drilling machine. The aperture of targets can be selected with a diameter of 2\farcs2, 2\farcs7, 3\farcs3, or 5\farcs3 according to the seeing. To avoid the blending of spectra, the Declination separation of targets should be larger than 5\farcs0. Within the FoV, 20 -- 50 apertures can typically be drilled in a single mask, with a maximum of one hundred apertures in ideal cases. The grisms of MOS are the same as those of BFOSC.

The observing procedures of MOS are as follows. Beside the bias, flats and lamp spectra, the flat lamps are imaged without the grism in place to illuminate the masks and provide the identification of targets. A direct image of the target field is also obtained to be used as comparisons. For the observations of standard stars, an extra mask is designed with a series of apertures with different diameters and locations, which can cover the whole wavelength range of the targets in any mask. One standard star may be observed totally three times with different locations, corresponding the wavelength range from blue to red.

 Our observations were carried out on the nights of 2012 September 15-17, using four multi-object masks with apertures of 2\farcs7 in diameter. Two 1800s exposures were secured for each field using the G6 and G8 grisms, providing spectra covering the $\sim$ 3000$-$5100 \AA\ and $\sim$ 6000$-$8100 \AA\ wavelength ranges, respectively (the coverage depends on the spatial distribution of the targets). One 1800s exposure offset the galaxy was made as the sky background. The seeing during the observations was $\sim$1\farcs5. The G6 grim allowed us to cover the region around [O\,{\sc ii}]\thinspace $\lambda$3727, H$\beta$, [O\,{\sc iii}]\thinspace $\lambda$4959, [O\,{\sc iii}]\thinspace $\lambda$5007 lines with 8.0 \AA\ Full Width Half Maxim spectral resolution, and G8 grism covers  H$\alpha$, [N\,{\sc ii}]\thinspace $\lambda$6583 and [S\,{\sc ii}]\thinspace $\lambda\lambda$6717, 6731 lines with a spectral resolution of 6.0 \AA.

\subsection{Data Reduction And Emission Line Fluxes}
The raw data were reduced with standard IRAF\footnote{IRAF is the Image Reduction and Analysis Facility written and supported by the IRAF programming group at the National Optical Astronomy Observatories (NOAO) in Tucson, Arizona which is operated by AURA, Inc. under cooperative agreement with the National Science Foundation} routines and some customized IRAF scripts. The reduction of spectra included bias subtraction, flat-field correction, cosmic ray removal, spectral extraction, wavelength and flux calibration. Different from normal long-slit spectra, wavelength ranges of multi-object masks changed with the spatial distribution of targets. We outline the key procedures here. First, the spectra from one mask were extracted after the raw two-dimensional images were bias-subtracted and trimmed. The spectral flats were used to define the trace for all apertures, and wavelength calibration was obtained using Iron Argon (FeAr) arc lamp exposures taken throughout each night. Then, the flat spectra in the same mask were normalized to their combined spectrum and were used to remove variations in response and illumination. For each night, at least three exposures of standard star HZ 4 were used for flux calibration, and each exposure was located in different position of one standard-star mask to make sure the wavelengths could cover all of the ranges of our targets. 

\begin{figure}[!htbp]
\center
\includegraphics[angle=0,scale=0.75]{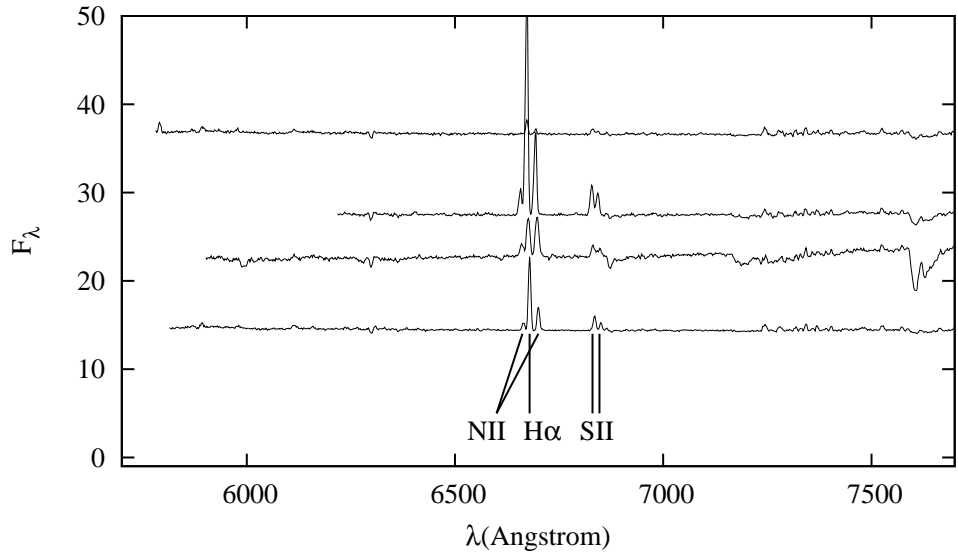}
\caption{Representative MOS spectra of H\,{\sc ii} regions in Arp 86. The spectra shown are from the same mask and observed with G8 grism. In order to show the spectra orderly, they are arranged artificially in the plot and the fluxes are in units of 10$^{-16}$ erg s$^{-1}$ cm$^{-2}$ \AA$^{-1}$.
\label{fig2}}
\end{figure}

The spectral intensities with observed wavelengths were corrected for the Milky Way extinction, the \citet{Fitzpatrick1999} extinction curve and the extinction A(V)=0.273 in the V band from \citet{Schlafly2011} were used. Figure~\ref{fig2} illustrates a sample of the resulting spectra extracted from one mask. The root mean squared (rms) noise of the continuum in the final corrected spectra is $\sim$ 1.0 $\times$ 10$^{-17}$ erg s$^{-1}$ cm$^{-2}$ \AA$^{-1}$. The spectrophotometric accuracy allows us to detect emission lines as weak as $\sim$ 10$^{-16}$ erg s$^{-1}$ cm$^{-2}$. Emission line fluxes were measured from the individual spectra by simultaneously fitting Gaussian profiles to the [O\,{\sc ii}]\thinspace$\lambda$3727, H$\beta$, [O\,{\sc iii}]\,$\lambda$5007, as well as H$\alpha$, [N\,{\sc ii}]\,$\lambda$6583 and [S\,{\sc ii}]\,$\lambda\lambda$6717, 6731 lines with the SPLOT program in IRAF. Most emission lines were detected with signal-to-noise ratios of S/N $>$ 5. For several H\,{\sc ii} regions, some lines in the blue part of the spectral range were below the detection limit.

\section{Analyses}
\label{sec3}
\subsection{Dust Extinction}
\label{sec3.1}
The dust extinction was estimated in order to map the dust distribution in galaxies, to correct interstellar reddening for emission lines, and to achieve extinction-free SFRs using intrinsic H$\alpha$ luminosities. H$\alpha$/H$\beta$ ratio was used to derive the value of the interstellar reddening from the Balmer decrement, because the relative intensities of the Balmer lines are nearly independent of both density and temperature \citep{Dominguez2013}. The usual de-reddening procedure is to derive the logarithmic extinction at H$\beta$, c($\beta$), and it can be measured in the following formula \citep{Stasinska2002}:

\begin{equation}
	c(H\beta) = \frac{R_{H\beta}}{R_{H\beta} - R_{H\alpha}} log \bigg[\frac{(H{\alpha}/H{\beta})_{obs}}{(H{\alpha}/H{\beta})_{int}}\bigg],
\end{equation}

where the subscript `obs' and `int' are the observed and corrected flux intensity ratios, respectively. We assumed the applicability of the \citet{Calzetti2000} extinction law, parametrized by $A_V = 4.0E(B-V) (R_V \equiv A_V /E(B-V) =4.0, R_{H\alpha} = 3.32, R_{H\beta} = 4.60)$. An intrinsic H$\alpha$/H$\beta$ ratio of 2.86 was assumed from Case B recombination at $10^4$K \citep{Osterbrock1989}.

\begin{figure}[!htbp]
\center
\includegraphics[angle=0,scale=0.6]{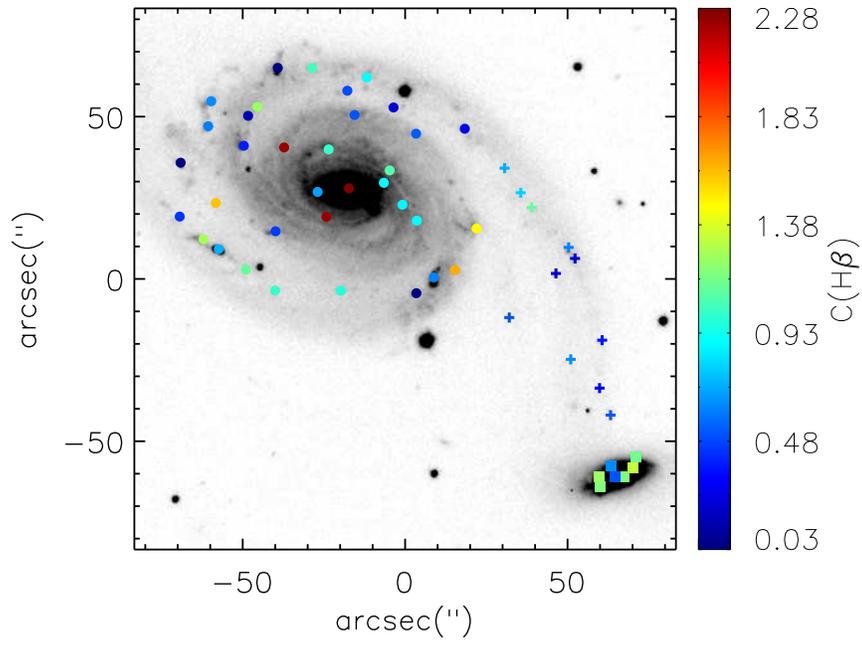}
\caption{The dust extinction c(H$\beta$) distribution across Arp 86. The symbols are same as in Figure~\ref{fig1}, where circles are from NGC 7753, squares from NGC 7752, and pluses from the bridge. 
\label{fig3}}
\end{figure}

Figure~\ref{fig3} shows the distribution of c(H$\beta$) across the disks and bridge of the galaxy pair. We found that the nucleus of NGC 7753 has the highest dust extinction, and the extinction in the bulge region of this galaxy is also higher than in other regions. In most regions of NGC 7552, the extinction is c(H$\beta$)$\sim$1.0, similar to that in the bulge of NGC 7753. However, the bridge has much lower extinction c(H$\beta$)$\sim$0.4 than those of the disks, which may indicate the low dust and gas densities in this structure.  

We corrected interstellar reddening for emission lines using the extinction obtained above. In a few cases ($\sim$ 1/5 of the sample) where H$\beta$ was too faint to be measured reliably, the value from one or several nearby sources with similar location was used. The extinction-corrected fluxes for the [O\,{\sc ii}]\thinspace $\lambda$3727, [O\,{\sc iii}]\thinspace$\lambda$5007, H$\alpha$, [N\,{\sc ii}]\thinspace $\lambda$6583 and [S\,{\sc ii}]\thinspace $\lambda\lambda$6717, 6731 emission lines are summarized in Table~\ref{table1}.

\subsection{Excitation Properties}
\begin{figure}[!htbp]
\center
\includegraphics[angle=0,scale=0.6]{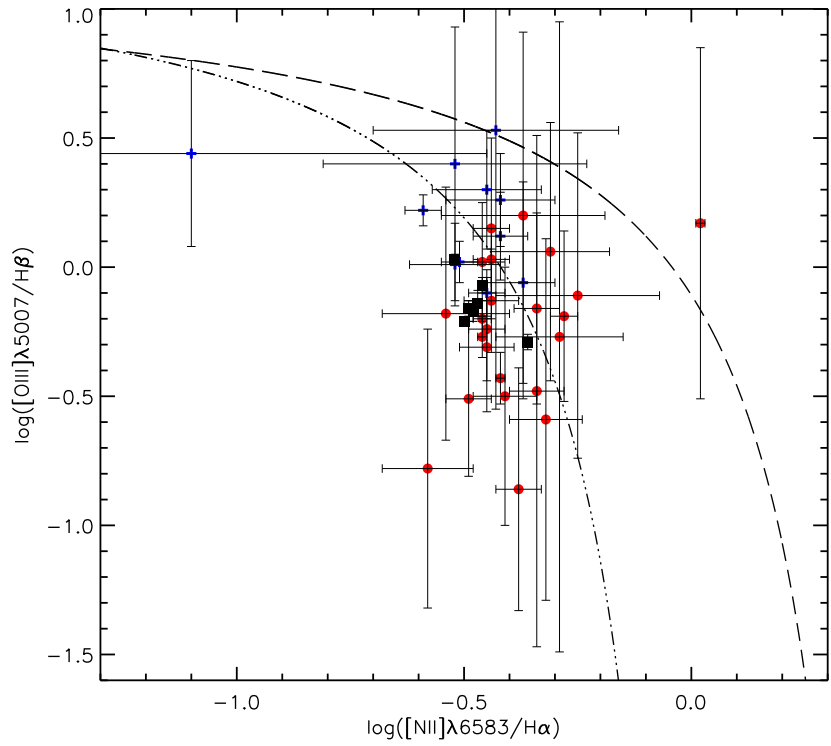}
\caption{Classical BPT diagram for the H\,{\sc ii} regions with the demarcation between star-forming regions and active galactic nuclei (AGN) from \citet{Kewley2001} (the long-dashed line) and \citet{Kauffmann2003} (the dot-dashed line). The symbols are same as in Figure~\ref{fig1}.
\label{fig4}}
\end{figure}

We inspected the excitation properties of our H\,{\sc ii} region sample by plotting the [O\,{\sc iii}]\thinspace $\lambda$5007/H$\beta$ line ratio against [N\,{\sc ii}]\thinspace $\lambda$6583/H$\alpha$ diagnostic diagrams \citep[BPT,][]{BPT1981}, as shown in Figure~\ref{fig4}. The diagram includes two demarcation curves, i.e., the maximum starburst line from \citet{Kewley2001} and the pure star forming line from \citet{Kauffmann2003}.

It can be seen that most of the regions are located under the maximum starburst line, and the sources in NGC 7752 are in the pure star forming region indicating that star formation is the dominant ionising source. The nucleus of NGC 7753 is the red point above the maximum starburst line, showing the AGN activity there. Some regions from the bridge and NGC 7753 lying in the composite region may be ionised by star formation, but also can be linked to the presence of shocks, AGB stars and AGN photoionization.

\subsection{Abundances Estimates}
In order to quantify the distribution of the metallicities in this interacting system the abundance diagnostic N2 = log([N\,{\sc ii}]\thinspace $\lambda$6583/H$\alpha$) was adopted to derive oxygen abundances in H\,{\sc ii} regions. The N2 index was studied by several groups \citep[e.g.,][]{Storchi1994, Raimann2000, Denicolo2002, Liang2006, Marino2013} and was confirmed to be a very useful indicator to calculate the oxygen abundance because of the close wavelength positions of these lines. We used the calibration from \citet{pp04}: $12 + log(O/H) = 8.9 + 0.57 \times N2$.

\begin{figure}[!htbp]
\center
\includegraphics[angle=0,scale=0.6]{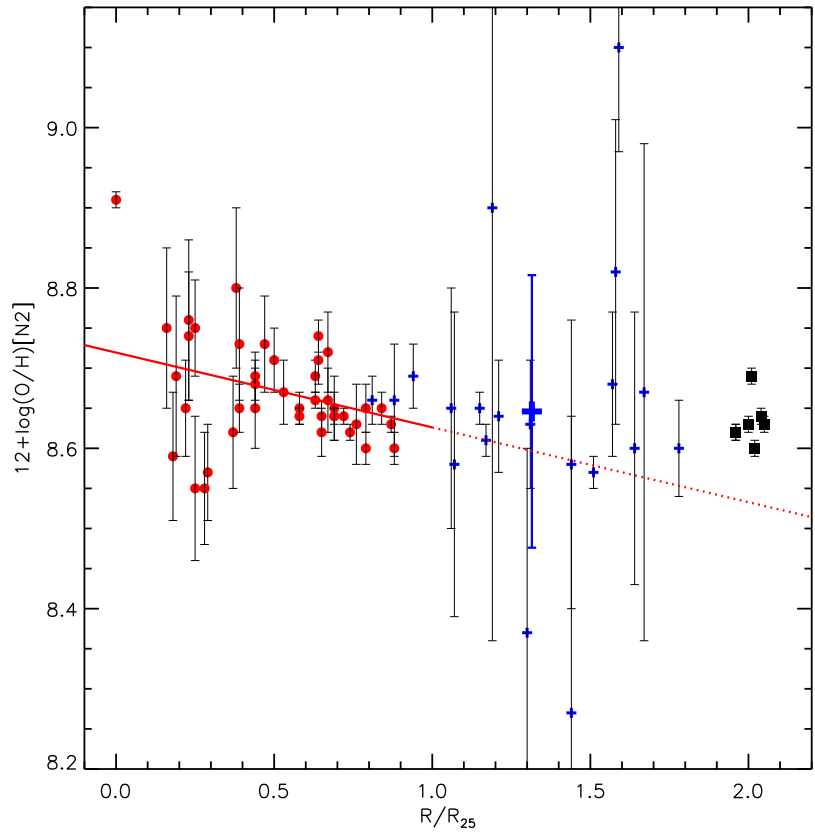}
\caption{Radial distribution of the oxygen abundance obtained from the N2 diagnostic. The red solid line is the linear least-squares fit of the H\,{\sc ii} regions in NGC 7753 (red circles), the red point line is the extension of the former. The symbols are same as in Figure~\ref{fig1}. The mean abundance in the tidal bridge is marked with the big plus with blue error bar. 
\label{fig5}}
\end{figure}

The resulting oxygen abundance is presented in Figure~\ref{fig5}. The center is located at the nucleus of NGC 7753, and the radius is normalized by the isophotal radii $R_{25}$ of NGC 7753. As the figure shows, most sources share the similar O/H ratio from 8.5 to 8.8. The nucleus of NGC 7753 is higher than all the regions in its disk due to the presence of AGN, which may act to bias the metallicity to higher value due to their enhanced N2 index. The median metallicity of NGC 7752 is found to be $\sim$ 8.63, similar to most regions in NGC 7753. The H\,{\sc ii} regions in the bridge have very low flux densities and low S/N, thus they have large dispersion in the abundance mainly due to the large errors, and their average abundance is $12 + log(O/H) \sim 8.65$.

Based on the abundances within the radii of R25, we made the linear fit of the metallicity as a function of radius to derive the radial gradient. The function is parameterized as $12 + log(O/H) = a + b (R/R_{25})$. The resulting linear least-squares fit is represented by the red line in Figure~\ref{fig5}, and can be expressed as $12 + log(O/H) = 8.72(\pm 0.02) - 0.09(\pm 0.04) R/R_{25}$. The slope of the abundance gradient is 0.09 dex $R_{25}^{-1}$, approximately -0.005 dex $kpc^{-1}$, is a little steeper than the outside regions and NGC 7752, which are located above the red-fit line.

\subsection{Star Formation Activities}
\label{STAR FORMATION ACTIVITIES}
Active star formation is the striking feature in interacting systems. As to Arp 86, we measured the SFR in each individual galaxies and H\,{\sc ii} regions. The total SFRs were calculated using the mid-infrared luminosities from \citet{Smith2007} and the relation of \citet{Wu2005a} because the fluxes in mid-infrared wavelength can be used as a proxy for star formation \citep{Wu2005a, Zhu2008}. The total SFRs are 8.6, 4.7 and 0.5 $M_{\odot}yr^{-1}$ for NGC 7753, NGC 7752 and the bridge region, respectively. The mean surface densities of SFRs ($\Sigma_{SFR}$) are 0.009, 0.081 and 0.003 $M_{\odot}yr^{-1}kpc^{-1}$ for NGC 7753, NGC 7752 and the bridge region, respectively. The $\Sigma_{SFR}$ in the companion galaxy is one order of magnitude higher than that in the main galaxy. The tidal bridge has the lowest SFR and $\Sigma_{SFR}$.

The SFRs for H\,{\sc ii} regions were calculated using the dust-corrected H$\alpha$ luminosity and the relation of \citet{Kennicutt1998}:
\begin{equation}
SFR_{H\alpha}(M_{\odot}yr^{-1})=7.9\times 10^{-42}[L(H\alpha)](erg~s^{-1}) \label{eq_SFR},
\end{equation}
which is calibrated using a Salpeter IMF with stellar masses in the range 0.1--100 M$_{\odot}$, and stellar population models with solar abundances. 

\begin{figure}[!htbp]
\center
\includegraphics[angle=0,width=0.45\textwidth]{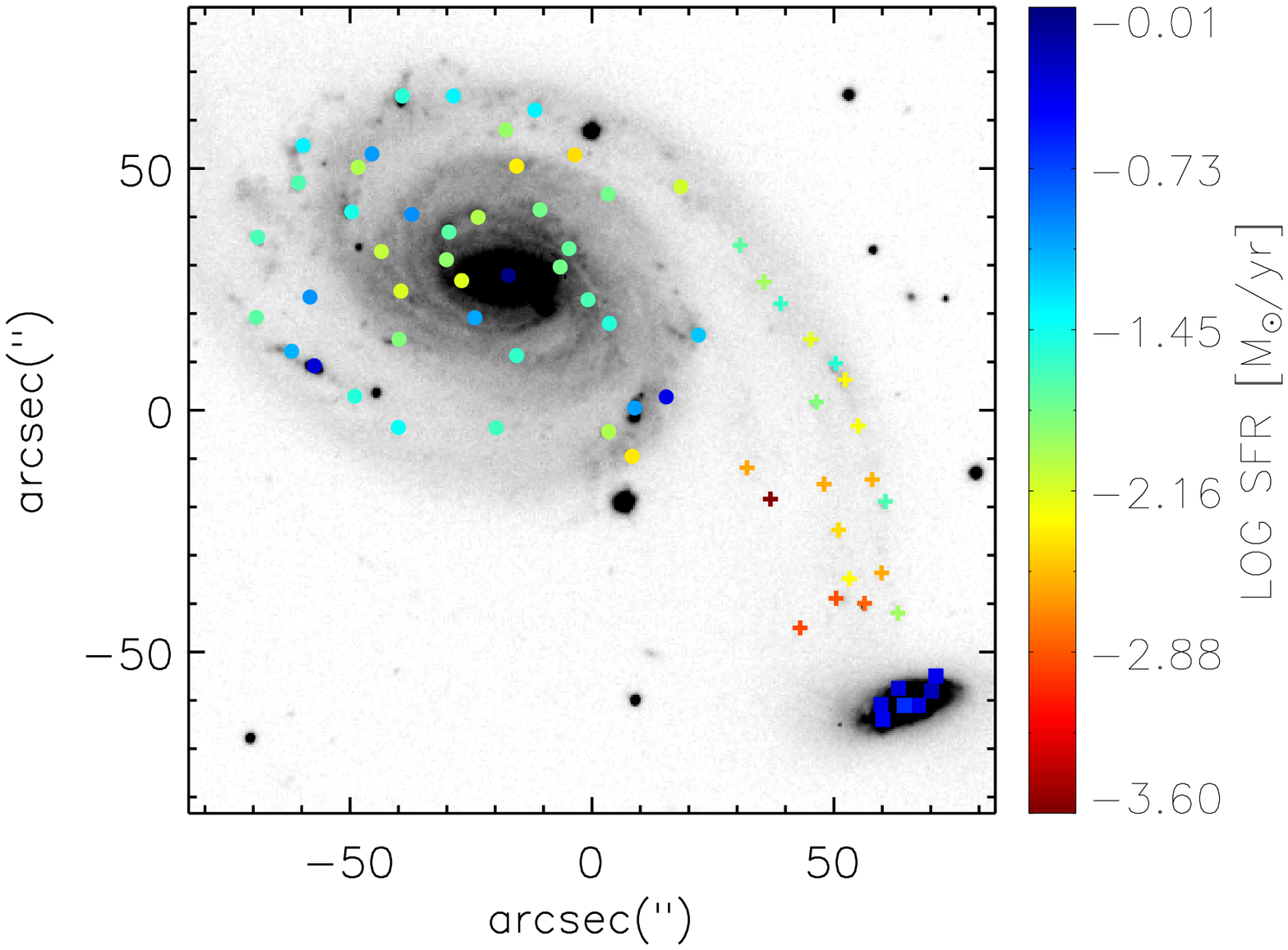}
\includegraphics[angle=0,width=0.45\textwidth]{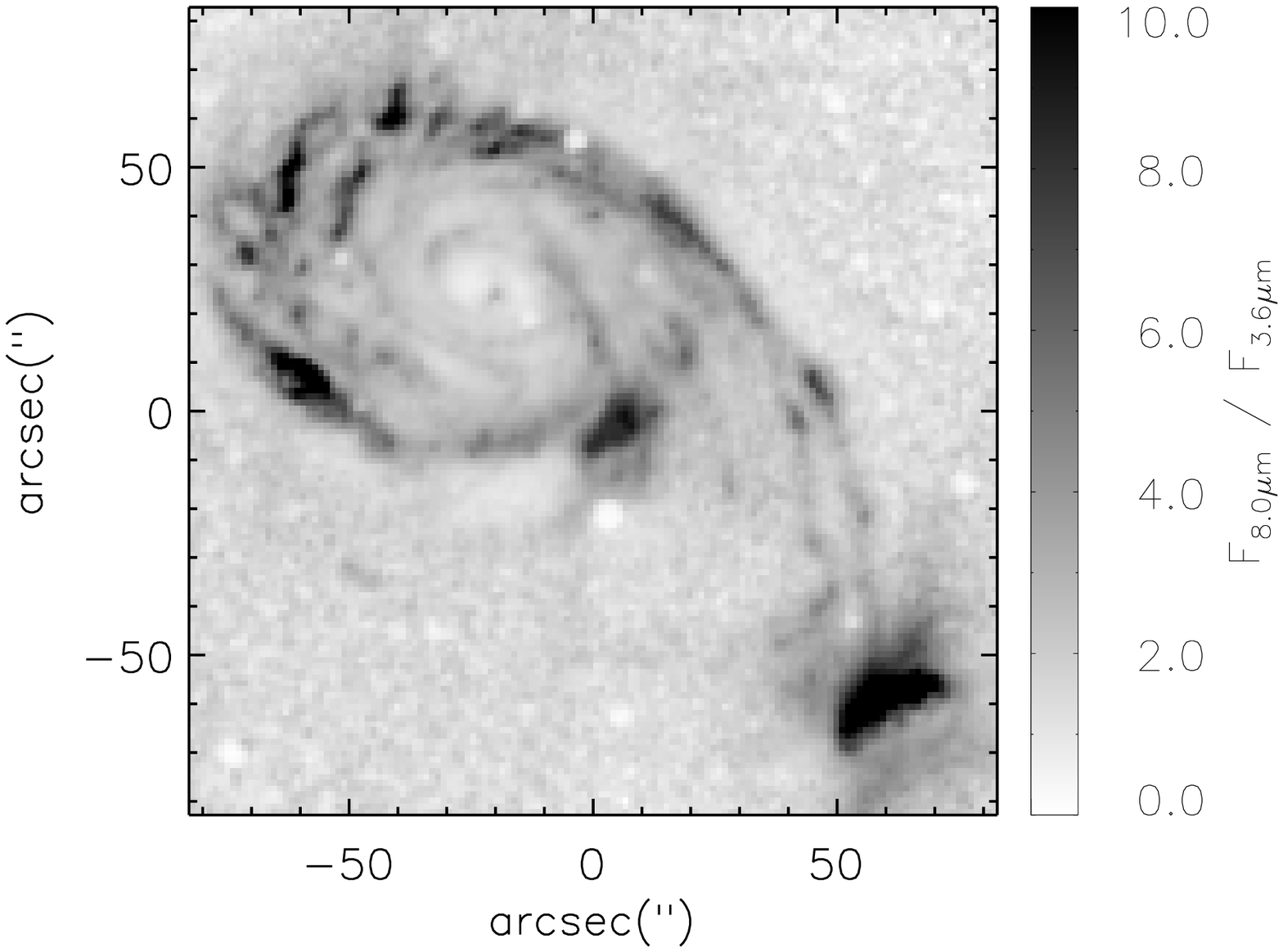}
\caption{Distribution of SFRs ({\it left}) and SSFRs ({\it right}) in Arp 86. In the left penal, the symbols are same as in Figure~\ref{fig1} with the colors corresponding to the measured SFRs. In the right penal, SSFRs are indicated by the ratio of the 8.0 $\mu$m and 3.6 $\mu$m images of Arp 86.
\label{fig6}}
\end{figure}

Figure~\ref{fig6} shows the detailed distribution of star formation activities in Arp 86. The left figure plots the SFRs with the background of optical image. We found that the companion galaxy NGC 7752 is currently experiencing a galaxy-wide starburst with log(SFR $M_{\odot}yr^{-1}$) = -0.6 -- -0.3 and $\Sigma_{SFR}$ = 0.5 -- 1.2 $M_{\odot}yr^{-1}kpc^{-1}$ for their H\,{\sc ii} regions. Most of the H\,{\sc ii} regions in NGC 7753 have log(SFR $M_{\odot}yr^{-1}$ ) lower than -1.0 except for the nucleus and two star-forming sources, which have similar SFRs to NGC 7752. However, the SFR in the nucleus of NGC 7753 may be overestimated due to the effect of the AGN. The star formation in the bridge is of log(SFR $M_{\odot}yr^{-1}$) $\sim$ -2.5, and the lowest $\Sigma_{SFR}$ in the interacting system as experted.

In order to further explore the star formation activities in Arp 86, we presented the ratio map of its 8.0 $\mu$m and 3.6 $\mu$m images in the right penal of Figure~\ref{fig6}. The two infrared images were taken from \citet{Smith2007} and observed by {\it Spitzer Space Telescope} \citep{Werner04}. Because the 3.6 $\mu$m luminosity provides an approximate measure of the stellar mass in the galaxy \citep{Zhu10} and the 8.0 $\mu$m luminosity can be used to trace star formation activities in galaxies, the ratio of the two images can be taken as the specific star formation rate (SSFR). From the ratio map, we found that the overall distribution of SSFRs in Arp 86 is similar to that of SFRs, where NGC 7752 has the highest SSFRs in the whole system and two regions lying in the southern arm of NGC 7753 also have obviously high SSFRs. In contrast, the SSFR in the bulge region of NGC 7753 is very low although the high SFR there.

\subsection{Stellar Population}
We used the spectral synthesis code STARLIGHT \citep{Cid2005, Mateus2006,
Asari2007, Cid2009} to derive the stellar populations of the target regions in Arp 86. This code fits an observed spectrum $O_\lambda$ with a model $M_\lambda$ which adds up a number of spectral components from a pre-defined set of base spectra. The  base spectra consist of simple stellar populations (SSPs) from \citet{BC03}, which have 25 ages from 0.001 to 18 Gyr and six metallicities from 0.005 to 2.5 $Z_{\odot}$, summing up 150 SSPs. 

\begin{figure}[!htbp]
\center
\includegraphics[angle=0,scale=0.6]{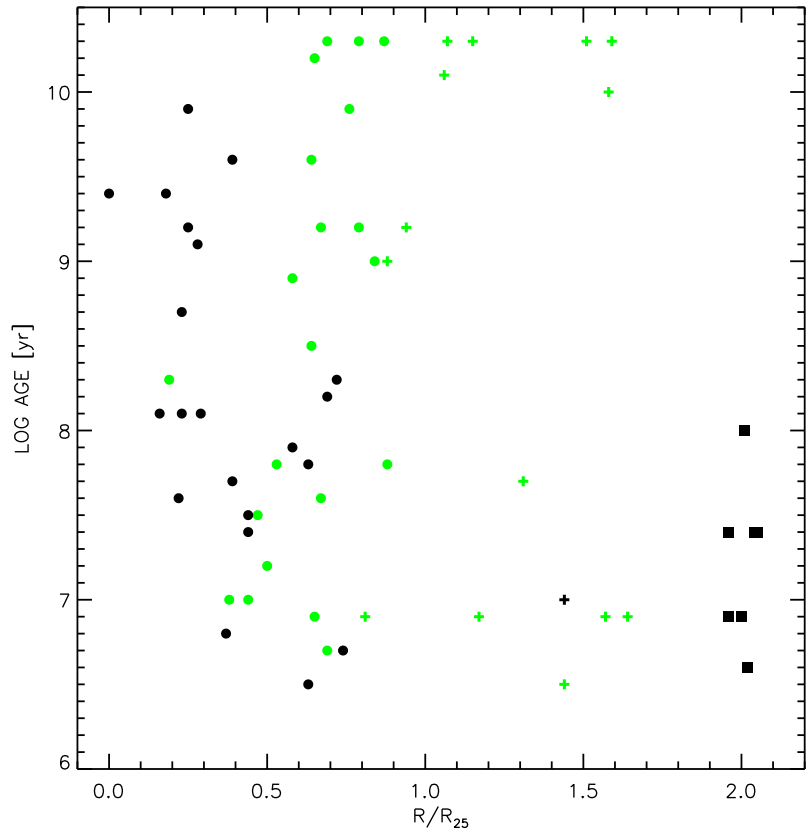}
\caption{Distribution of the stellar ages in Arp 86. The ages are weighted by stellar luminosity \citep{Cid2005} and derived from spectral synthesis fitting. The {\it black} symbols are the spectra with S/N $\geq $ 3.0 (measured at 4730 -- 4780 \AA), and the {\it green} ones are the spectra with S/N $<$ 3.0.
\label{fig7}}
\end{figure}

The results of the spectral synthesis fitting are presented in Figure~\ref{fig7}, where the distribution of stellar ages weighted by the light fraction is plotted. Because the spectra with low S/N may result in large errors, we divided the spectra into two class based on S/N measured at 4730 -- 4780 \AA~ in Figure~\ref{fig7}: S/N $\geq $ 3.0 (in black color) and S/N $<$ 3.0 (in green color). We mainly consider the spectra with S/N $\geq $ 3.0. It can be seen from Figure~\ref{fig7} that the ages of target regions in NGC 7752 are $\sim$ 10 -- 100 Myr, similar to the tidal bridge and some arm regions in NGC 7753. The ages of the bulge region in NGC 7753 are $\sim$ 1 -- 10 Gyr, according with the low SSFR mentioned in Sec~\ref{STAR FORMATION ACTIVITIES}.

\section{Discussion}
\label{sec4}
In the previous section we mainly illustrated the properties of abundances and star formation activities in the interacting galaxy pair Arp 86. The slope of the abundance gradient in the inner disk of NGC 7753 is less than 0.1 dex $R_{25}^{-1}$ along with a much shallower slope in the outer disk and tidal bridge. H\,{\sc ii} regions in NGC 7752 have higher SFRs than those in NGC 7753, one order of magnitude higher in common. The stellar populations in NGC 7752 are in the age of $\sim$ 10 -- 100 Myr. These evidences indicate that more active star formation is ongoing in the companion galaxy than in the main galaxy. These results provide solid evidence that galaxy interactions play an important role in modifying the metallicity properties and star formation activities of galaxies.

Simulations indicate that tidal forces during encounters can cause large-scale gas inflows in galaxies \citep{Barnes1996}, which can flatten the metallicity distributions \citep{Rupke2010}. \citet{Kewley2006} proposed a scenario in which galaxy interactions drive large gas inflows toward the central regions, less enriched gas is carried from the outskirts of the galaxy into the central regions, disrupting metallicity gradients, and diluting central metallicities. Consistently with these models, our results from Figure~\ref{fig5} show a very flatting abundance gradient with the slope of 0.09 $\pm$ 0.04 dex $R_{25}^{-1}$ in the inner disk of the main galaxy, which is corresponding to $\sim$ 0.005 dex kpc$^{-1}$. The slope is one order of magnitude lower than the typical values of 0.03--0.10 dex kpc$^{-1}$ in normal spiral galaxies, such as NGC 628 \citep{Gusev2013}, NGC 3621 \citep{Bresolin2012}, M33 \citep{Bresolin2011}, M101 \citep{Lin2013}. Their observational results provide similar evidences. \citet{Kewley2010} systematically studied a sample of close galaxy pairs and found that the mean gradient in their galaxy pairs is -0.25 dex $R_{25}^{-1}$ compared with a mean gradient of -0.67 dex $R_{25}^{-1}$ for the isolated spiral galaxies, and proved galaxy pairs have flatter metallicity gradients due to large tidal gas inflows in galaxy interactions and mergers.

Interaction-induced star formation has also been investigated by observations and models. In many sample and case studies, ultraviolet, H$\alpha$ and infrared images show enhanced star formation activities in interacting systems \citep[e.g.,][]{Cao2007,Smith2007}. The simulations of Cox et al. (2008) found merger-driven star formation is a strong function of merger mass ratio, and the less massive companion is expected to be more susceptible to tidal forces and may have a larger enhancement of star formation. In our results, Figure~\ref{fig6} and \ref{fig7} show the same trend, where the SFRs of H\,{\sc ii} regions are one order of magnitude higher in NGC 7752 than in the main galaxy NGC 7753. Such evidences are also found in NGC 7771+NGC 7770 interacting system \citep{Alonso2012} and statistics based on Sloan Digital Sky Survey Data \citep{Woods2007}.

Beside NGC 7753 and 7752, the galaxy 2MASX J23470758+2926531 may be a part of the Arp 86 system. This galaxy is visible in the GALEX UV images \citep{Smith2010}, along with many optical images. \citet{Sengupta2009} estimated the stellar mass and H\,{\sc ii} mass of this galaxy, and found they are 2.9 $\times 10^9 M_{\odot}$ and 4.5 $\times 10^8 M_{\odot}$, respectively. HI column density image shows an connection between NGC 7752 and 2MASX J23470758+2926531 \citep{Sengupta2009}. We also obtained the spectra of 2MASX J23470758+2926531, and found its redshift $z = 0.0166 \pm 0.0003$, similar to Arp 86 (z $\sim$ 0.0162). It is located at the composite region in the BPT diagram, the $\Sigma_{SFR}$ is $\sim$ 0.027 $M_{\odot}yr^{-1}kpc^{-1}$, and the oxygen abundance is similar to most of regions in NGC 7753, which prove that 2MASX J23470758+2926531 is an companion of NGC 7753.

\section{SUMMARY}
\label{sec5}
Optical MOS spectra of 69 H$\alpha$ regions located in the interacting system Arp86 have been analyzed. The oxygen abundances, SFRs and stellar populations have been derived. 
We have found that the extinction in NGC 7552 and the bulge region of NGC 7753 is c(H$\beta$)$\sim$1.0, higher than that in the bridge and out disks of NGC 7753 (c(H$\beta$) $\sim$ 0.4 -- 0.8). The BPT diagram indicates the pure star formation activity in NGC 7752 and AGN activity in the nucleus of NGC 7753. 

The oxygen abundances estimated from the N2 metallicity indicator yield the similar O/H ratio from 8.5 to 8.8 in most H\,{\sc ii} regions. The slope of the abundance gradient in the inner disk of NGC 7753 is less than 0.1 dex $R_{25}^{-1}$ along with a much shallower slope in the outer disk and tidal bridge, indicating that tidal forces during encounters can flatten the metallicity distributions. 

The total SFRs are 8.6, 4.7 and 0.5 $M_{\odot}yr^{-1}$ for NGC 7753, NGC 7752 and the bridge region, respectively. The companion galaxy NGC 7752 is currently experiencing a galaxy-wide starburst with log(SFR $M_{\odot}yr^{-1}$) = -0.6 -- -0.3 and $\Sigma_{SFR}$ = 0.5 -- 1.2 $M_{\odot}yr^{-1}kpc^{-1}$ for their H\,{\sc ii} regions. We also have found that the $\Sigma_{SFR}$ of H\,{\sc ii} regions are one order of magnitude higher in NGC 7752 than in the main galaxy NGC 7753, and the stellar population of NGC 7752 is also as young as 10 -- 100 Myr, predicting that the companion galaxy is more susceptible to the effects of the interaction than the primary.

We also have observed the spectra of the galaxy 2MASX J23470758+2926531, and have found that this galaxy has similar redshift to Arp 86, and its abundance and star formation properties are similar to NGC 7753, indicating it may be a part of the Arp 86 system.

\begin{acknowledgements}
We are grateful to the anonymous referee for thoughtful comments and insightful suggestions that helped to improve this paper. 
We thank the kind staff at the Xinglong 2.16m telescope for their support during the observations. This project is supported by the Chinese National Natural Science Foundation grands No. 11078017, 11303038, 11073032, 11373035, 11203034, 11203031 and 11303043, and by the National Basic Research Program of China (973 Program), No. 2014CB845704, 2013CB834902, and 2014CB845702.
We made extensive use of the NASA/IPAC Extragalactic Database (NED) which is operated by the Jet Propulsion Laboratory, California Institute of Technology, under contract with the National Aeronautics and Space Administration.
\end{acknowledgements}

\clearpage
\small
\setlength\LTleft{-0.6in}
\setlength\LTright{-1in plus 1 fill}
\begin{longtable}{p{0.1cm}cccccccccc}
	\caption{{\small Observed H{\sc ii} Region Sample}} \label{table1}
 \tabularnewline\hline
	ID & R.A. & Dec. & [O\,{\sc ii}] & [O\,{\sc iii}] &  H$\alpha$ & [N\,{\sc ii}] & [S\,{\sc ii}] & c(H$\beta$) & 12+log(O/H) & log SFR\\
	 & (J2000.0) & (J2000.0) & $\lambda$3727 & $\lambda$5007 &  & $\lambda$6583 & $\lambda\lambda$6717,6713 & &  & $M_{\odot}yr^{-1}$\\  \hline
 \endfirsthead
 \multicolumn{11}{c}%
{{\bfseries \tablename\ \thetable{} -- Continued}}\\
\hline
	ID & R.A. & Dec. & [O\,{\sc ii}] & [O\,{\sc iii}] &  H$\alpha$ & [N\,{\sc ii}] & [S\,{\sc ii}] & c(H$\beta$) & 12+log(O/H) & log SFR\\
	 & (J2000.0) & (J2000.0) & $\lambda$3727 & $\lambda$5007 &  & $\lambda$6583 & $\lambda\lambda$6717,6713 &  &  & $M_{\odot}yr^{-1}$\\  \hline
\endhead
1&23:47:05.68&29:29:37.68&187(13)&25.4(7.4) & 125(3) & 44.0(2.9) & 47.2(7.8) & 1.05 & 8.64 & -1.34\\
2 & 23:47:06.49 & 29:29:37.67 & 47.7(2.9) & 18.2(1.9) & 82.1(0.8) & 28.5(0.7) & 28.0(1.4) & 0.04 & 8.64 & -1.52\\
3 & 23:47:04.39 & 29:29:34.78 & 15.4(4.8) & 48.3(5.8) & 126(3) & 45.6(3.2) & 68.3(6.1) & 0.91 & 8.65 & -1.33\\
4 & 23:47:04.85 & 29:29:30.68 & 24.8(5.5) & ... & 32.9(1.6) & 15.2(1.5) & 22.8(2.2) & 0.49 & 8.71 & -1.92\\
5 & 23:47:08.06 & 29:29:27.40 & 53.1(3.3) & ... & 128(2) & 46.7(2.0) & 42.6(3.3) & 0.63 & 8.65 & -1.32\\
6 & 23:47:06.97 & 29:29:25.68 & 124(14) & ... & 257(5) & 110.8(5.0) & 76.4(10.7) & 1.24 & 8.69 & -1.02\\
7 & 23:47:03.76 & 29:29:25.52 & ... & ... & 11.0(0.9) & 5.41(0.82) & 13.7(1.3) & 0.22 & 8.72 & -2.39\\
8 & 23:47:04.68 & 29:29:23.19 & ... & ... & 12.5(1.8) & 8.41(2.16) & 6.57(2.89) & 0.51 & 8.80 & -2.34\\
9 & 23:47:07.19 & 29:29:22.90 & 12.4(0.9) & 6.45(4.67) & 26.8(0.9) & 12.4(0.9) & 14.6(2.3) & 0.14 & 8.71 & -2.00\\
10 & 23:47:08.13 & 29:29:19.69 & ... & ... & 59.6(1.9) & 17.6(0.9) & 50.0(1.6) & 0.61 & 8.60 & -1.66\\
11 & 23:47:02.08 & 29:29:18.99 & 22.2(3.6) & 1.90(1.93) & 20.9(1.1) & 9.99(1.34) & 23.4(2.1) & 0.29 & 8.72 & -2.11\\
12 & 23:47:03.23 & 29:29:17.46 & 34.8(2.9) & 2.14(2.05) & 43.7(1.9) & 18.1(1.3) & 20.7(3.1) & 0.53 & 8.68 & -1.79\\
13 & 23:47:04.31 & 29:29:14.17 & ... & ... & 44.0(3.0) & 24.2(3.9) & 34.1(6.9) & ... & 8.75 & -1.79\\
14 & 23:47:07.29 & 29:29:13.69 & ... & ... & 91.5(1.5) & 32.2(1.0) & 24.7(2.1) & 0.39 & 8.64 & -1.47\\
15 & 23:47:06.34 & 29:29:13.18 & 3540(266) & 119(53) & 285(32) & 140(25) & 159(31) & 2.23 & 8.73 & -0.98\\
16 & 23:47:05.29 & 29:29:12.59 & 31.3(53.6) & 7.13(4.29) & 25.8(2.8) & 14.4(4.3) & 33.9(4.5) & 1.02 & 8.76 & -1.69\\
17 & 23:47:05.75 & 29:29:09.54 & ... & ... & 55.2(2.7) & 13.2(3.9) & 29.1(3.9) & ... & 8.55 & -1.76\\
18 & 23:47:08.78 & 29:29:08.43 & 9.07(0.59) & ... & 62.3(0.8) & 21.0(0.7) & 16.8(1.3) & 0.03 & 8.63 & -1.64\\
19 & 23:47:01.14 & 29:29:06.87 & 36.3(3.9) & 24.8(3.2) & 53.3(1.8) & 20.4(2.0) & 35.3(5.3) & 0.71 & 8.66 & -1.71\\
20 & 23:47:03.85 & 29:29:06.14 & 169(33) & ... & 50.5(3.1) & 26.1(6.5) & 42.6(5.7) & 1.08 & 8.74 & -1.73\\
21 & 23:47:06.82 & 29:29:05.50 & ... & ... & 23.3(1.2) & 8.39(1.12) & 9.66(2.04) & ... & 8.65 & -2.06\\
22 & 23:47:05.79 & 29:29:03.81 & ... & ... & 34.2(1.8) & 12.6(2.2) & 14.9(5.4) & ... & 8.65 & -1.90\\
23 & 23:47:03.99 & 29:29:02.33 & 49.4(4.6) & 10.3(4.2) & 43.9(3.1) & 12.6(3.4) & 18.6(2.6) & 0.90 & 8.59 & -1.79\\
24 & 23:47:04.81 & 29:29:00.60 & 12569(1535) & ... & 2650(49) & 2790(43) & 897(40) & 2.28 & 8.91 & -0.01\\
25 & 23:47:05.55 & 29:28:59.50 & ... & ... & 18.5(2.3) & 10.1(2.9) & 9.91(2.49) & 0.69 & 8.75 & -2.17\\
26 & 23:47:00.76 & 29:28:59.29 & 42.7(5.1) & 19.4(2.7) & 30.2(2.1) & 11.4(2.4) & 20.6(2.6) & 0.78 & 8.66 & -1.95\\
27 & 23:47:06.51 & 29:28:57.31 & ... & ... & 18.8(1.4) & 6.07(1.19) & 10.4(2.3) & ... & 8.62 & -2.16\\
28 & 23:47:07.95 & 29:28:56.06 & 409(37) & 103(31) & 276(8) & 95.5(7.8) & 66.0(21.3) & 1.60 & 8.64 & -0.99\\
29 & 23:47:03.55 & 29:28:55.58 & 42.3(12.0) & 3.60(3.54) & 60.9(2.7) & 16.0(3.1) & 40.0(4.0) & 0.87 & 8.57 & -1.65\\
30 & 23:47:00.50 & 29:28:54.78 & 83.0(11.9) & 22.3(9.7) & 72.0(3.8) & 30.7(3.4) & 45.9(6.6) & 1.12 & 8.69 & -1.58\\
31 & 23:47:08.80 & 29:28:51.85 & 66.6(3.1) & ... & 53.2(1.3) & 16.0(1.1) & 22.6(3.1) & 0.44 & 8.60 & -1.71\\
32 & 23:47:05.34 & 29:28:51.81 & ... & 137(104) & 240(31) & 102(29) & 55.3(29.1) & 2.23 & 8.69 & -1.05\\
33 & 23:47:03.21 & 29:28:50.68 & 32.7(18.9) & 14.7(5.9) & 84.0(2.4) & 30.0(3.1) & 48.9(4.8) & 0.91 & 8.64 & -1.51\\
34 & 23:47:01.80 & 29:28:48.28 & 457(35) & 48.2(11.1) & 183(7) & 65.5(7.0) & 103(10) & 1.47 & 8.65 & -1.17\\
35 & 23:47:00.02 & 29:28:47.44 & ... & ... & 17.1(2.4) & 4.78(3.04) & 9.21(3.56) & ... & 8.58 & -2.20\\
36 & 23:47:06.54 & 29:28:47.34 & ... & ... & 39.2(1.5) & 16.6(1.6) & 19.0(2.8) & 0.46 & 8.69 & -1.84\\
37 & 23:47:08.24 & 29:28:45.00 & 469(62) & 107(35.7) & 214(6) & 77.4(5.5) & 118(16.4) & 1.27 & 8.65 & -1.10\\
38 & 23:47:04.68 & 29:28:44.01 & ... & ... & 73.1(4.8) & 17.7(3.8) & 34.6(5.6) & ... & 8.55 & -1.57\\
39 & 23:46:59.63 & 29:28:42.48 & 93.1(3.9) & 29.5(2.8) & 80.2(1.7) & 24.8(1.6) & 26.4(3.5) & 0.61 & 8.61 & -1.53\\
40 & 23:47:07.88 & 29:28:41.83 & 1110(14) & ... & 1320(4) & 419(3) & 253(4) & 0.71 & 8.62 & -0.31\\
41 & 23:46:59.48 & 29:28:39.07 & 24.4(3.4) & 9.62(2.79) & 13.7(0.8) & 4.86(1.10) & 8.15(2.00) & 0.22 & 8.64 & -2.30\\
42 & 23:47:07.24 & 29:28:35.56 & 259(17) & ... & 83.6(3.2) & 31.4(4.2) & 70.7(5.6) & 1.12 & 8.66 & -1.51\\
43 & 23:47:02.31 & 29:28:35.50 & 795(73) & 224(27) & 1180(13) & 407(8) & 300(17) & 1.65 & 8.64 & -0.36\\
44 & 23:46:59.93 & 29:28:34.70 & 27.9(1.4) & 10.7(1.2) & 38.2(0.9) & 13.7(0.8) & 14.4(1.7) & 0.21 & 8.66 & -1.85\\
45 & 23:47:02.81 & 29:28:33.20 & 128(12) & 33.7(5.9) & 255(2) & 97.4(1.8) & 80.3(3.0) & 0.62 & 8.66 & -1.03\\
46 & 23:46:59.27 & 29:28:29.54 & ... & ... & 15.1(1.4) & 5.00(1.23) & 10.6(1.6) & ... & 8.63 & -2.26\\
47 & 23:47:06.55 & 29:28:29.12 & 130(12) & 11.5(5.0) & 104(4) & 33.9(2.9) & 30.6(7.0) & 1.01 & 8.62 & -1.41\\
48 & 23:47:05.00 & 29:28:29.05 & 88.0(9.6) & 7.38(7.25) & 66.4(3.4) & 26.0(3.0) & 37.0(8.2) & 0.98 & 8.67 & -1.61\\
49 & 23:47:03.22 & 29:28:28.30 & 16.8(2.1) & 6.18(4.07) & 27.2(0.7) & 14.3(0.7) & 16.0(1.4) & 0.03 & 8.74 & -2.00\\
50 & 23:47:02.85 & 29:28:23.19 & ... & ... & 12.0(0.5) & 4.05(0.65) & 7.40(1.13) & ... & 8.63 & -2.35\\
51 & 23:47:01.03 & 29:28:20.85 & 15.0(6.4) & 8.40(9.18) & 6.99(1.40) & 2.59(1.09) & 11.6(2.1) & 0.50 & 8.65 & -2.59\\
52 & 23:46:59.05 & 29:28:18.43 & ... & ... & 7.63(1.17) & 2.10(1.17) & 9.65(2.38) & ... & 8.58 & -2.55\\
53 & 23:46:59.81 & 29:28:17.49 & ... & ... & 7.23(1.26) & 0.841(0.649) & 6.76(1.83) & ... & 8.37 & -2.57\\
54 & 23:47:00.66 & 29:28:14.39 & 35.1(6.7) & ... & 0.683(0.781) & 0.693(0.714) & 11.6(2.3) & ... & 8.90 & -3.60\\
55 & 23:46:58.84 & 29:28:13.90 & 95.6(9.5) & 35.4(1.8) & 61.3(1.3) & 15.9(1.1) & 16.9(2.1) & 0.38 & 8.57 & -1.65\\
56 & 23:46:59.58 & 29:28:07.99 & 63.0(7.9) & 10.3(2.4) & 10.5(1.5) & 0.829(1.130) & 21.6(2.5) & 0.64 & 8.27 & -2.41\\
57 & 23:46:58.90 & 29:27:59.14 & 5.50(3.03) & 6.38(5.14) & 7.15(1.30) & 2.15(1.03) & 11.3(1.7) & 0.29 & 8.60 & -2.58\\
58 & 23:46:59.41 & 29:27:57.92 & ... & ... & 13.5(1.3) & 5.56(1.38) & 13.5(2.3) & ... & 8.68 & -2.30\\
59 & 23:46:59.62 & 29:27:53.87 & ... & ... & 3.44(0.87) & 7.69(2.12) & 23.9(1.9) & ... & 9.10 & -2.90\\
60 & 23:46:59.17 & 29:27:52.85 & ... & ... & 3.99(1.65) & 1.60(1.32) & 11.8(1.7) & ... & 8.67 & -2.83\\
61 & 23:46:58.64 & 29:27:50.82 & 13.0(2.5) & 11.2(2.1) & 30.8(1.6) & 9.33(1.73) & 15.3(1.9) & 0.52 & 8.60 & -1.94\\
62 & 23:47:00.19 & 29:27:47.75 & ... & ... & 3.24(0.71) & 2.37(1.25) & 18.4(1.8) & ... & 8.82 & -2.92\\
63 & 23:46:58.04 & 29:27:37.83 & 1384(28) & 420(7) & 1110(7) & 331(6) & 252(8) & 1.15 & 8.60 & -0.39\\
64 & 23:46:58.63 & 29:27:35.26 & 1302(9) & 300(4) & 1380(4) & 439(3) & 291(6) & 0.63 & 8.62 & -0.29\\
65 & 23:46:58.11 & 29:27:34.74 & 2841(70) & 527(24) & 1730(11) & 595(8) & 526(15) & 1.33 & 8.64 & -0.19\\
66 & 23:46:58.91 & 29:27:31.93 & 1648(29) & 307(13) & 1240(8) & 404(6) & 339(12) & 1.25 & 8.62 & -0.34\\
67 & 23:46:58.32 & 29:27:31.74 & 2216(33) & 285(16) & 1190(9) & 398(5) & 324(9) & 1.19 & 8.63 & -0.36\\
68 & 23:46:58.54 & 29:27:31.73 & 344(9) & 116(5) & 650(2) & 283(2) & 297(5) & 0.51 & 8.69 & -0.62\\
69 & 23:46:58.88 & 29:27:28.81 & 1523(36) & 274(17) & 1060(9) & 360(7) & 355(12) & 1.19 & 8.63 & -0.41\\
70 & 23:47:07.56 & 29:26:54.13 & 60.9(7.3) & 17.6(7.0) & 33.8(1.4) & 14.2(1.8) & 34.7(4.2) & 0.52 & 8.69 & -1.90\\  
\noalign{\smallskip}\hline
\end{longtable}
\tablecomments{\textwidth}{
	Line fluxes are in the unit of $10^{-16}$ erg s$^{-1}$ cm$^{-2}$, after correcting for reddening and extinction.
	The values in brackets are the flux errors of the emission lines.
}

\end{document}